\documentclass[
 aip,
 sd,%
amsart,amsmath,amssymb,
 reprint,%
superscriptaddress,
]{revtex4-1}

\usepackage{dcolumn}
\usepackage{bm}
\usepackage[usenames, dvipsnames]{color}
\usepackage{multirow}
\usepackage{float}
\usepackage[pdftex]{graphicx}
\usepackage{url}
\usepackage{natbib,hyperref}

\begin{document}

\preprint{AIP/123-QED}

\title{On the influence of the intermolecular potential on the wetting properties of water on silica surfaces\footnote{Error!}}
\author{E. Pafong}
\affiliation{Institut f\"ur Festk\"orperphysik, Technische Universit\"at 
Darmstadt, Hochschulstr. 6, 64289 Darmstadt, Germany}
\thanks{priscel@fkp.tu-darmstadt.de}
\author{J. Geske}
\affiliation{Institut f\"ur Festk\"orperphysik, Technische Universit\"at 
Darmstadt, Hochschulstr. 6, 64289 Darmstadt, Germany}
 \author{B. Drossel}
\affiliation{Institut f\"ur Festk\"orperphysik, Technische Universit\"at 
Darmstadt, Hochschulstr. 6, 64289 Darmstadt, Germany}
\date{\today}

\begin{abstract}
We study the wetting properties of water on silica surfaces using molecular
dynamics (MD) simulations. To describe the intermolecular interaction between
water and silica atoms, two types of interaction potential models are used: the
standard Br{\'o}dka and Zerda (BZ) model, and the Gulmen and Thompson (GT)
model. We perform an in-depth analysis of the influence of the choice of the
potential on the arrangement of the water molecules in partially filled pores
and on top of silica slabs. We find that at moderate pore filling ratios,
the GT silica surface is completely wetted by water molecules, which
agrees well with experimental findings, while the commonly used BZ
surface is less hydrophilic and is only partially wetted. We interpret
our simulation results using an analytical calculation of the phase diagram of
water in partially filled pores. Moreover, an evaluation of the contact angle
of the water droplet on top of the silica slab reveals that the interaction
becomes more hydrophilic with increasing slab thickness and saturates around
$2.5-3$\,nm, in agreement with the experimentally found value. Our analysis
also shows that the hydroaffinity of the surface is mainly determined by the
electrostatic interaction, but that the van der Waals interaction nevertheless
is strong enough that it can turn a hydrophobic surface into a hydrophilic surface.  
\end{abstract}

\maketitle
\section{Introduction}
Water is an essential material in our everyday life and is the most used
solvent for chemical and biological reactions. Water molecules are highly
polar, forming hydrogen-bonded networks and sharing hydrogen bonds with other
molecules. In particular, water confined within nanoscale geometries of
hydrophilic surfaces is subject to two competing interactions: the hydrophilic interactions between water molecules and those between water and surface molecules.\\ \\
One of the standard systems for studying hydrophilic interactions with water are silica nanopores such as sols-gels, \cite{Sco_Four} mesoporous silica (MCM-41), \cite{Grun_Bunt, Taka_Yama, Fara_Chen, Smir_Kuro, Taka_Belli, Kit_Yama} Vycor-glasses, \cite{Ross_Benh, Bru_So, Thom_Skip} controlled pore glasses (CPG). \cite{Vya_Buntk,  Vya_Bunt, Al_Gub, Fog_Thom, Thomp}
Water confined in silica nanopores or near silica flat surfaces is a topic
which has attracted considerable attention,\cite{Harr_Vogel, Fog_Pan,
Brov_Pas, Gallo_Spohr, Mill_La,  Ren_Szy, Zhang_Sing} mainly because of the
relevance of the water-silica interaction in understanding water transport in
porous rocks, \cite{Berk} nanofluidic devices, \cite{Sto_Ajd} heteregeneous
catalysis in mesoporous materials, \cite{Cor_Vel, Thomp} and permeation through membrane channels. \cite{Ber_Roux} 
Experimental investigations of water in silica nanopores have been carried out
using NMR spectroscopy,\cite{Grun_Bunt, Vya_Bunt} X-ray and neutron
diffraction,\cite{Bru_So, Ri_So, Thom_Skip, Jel_Dor, Smir_Kuro, Taka_Belli}
quasi-elastic neurton scattering, \cite{Taka_Yama, Fara_Chen} Small-angle
neutron scattering (SANS),\cite{Ross_Benh} and optical Kerr-effect spectroscopy,\cite{Sco_Four} showing that the dynamics of water in such pores is slow in
comparison to the dynamics of bulk water. This originates from the strong
binding or trapping of water molecules by silica surfaces as found by
experimental measurements conducted with MCM-41 as well as CPG pores
\cite{Grun_Bunt, Vya_Bunt} resulting in a complete coverage of the pore
surface at even moderate hydration levels. Accordingly, the intermolecular
interactions between water and silica surfaces in MD simulations should be set
up such that the experimental results are reproduced. \\ \\ 
The influence of the filling ratios on the wetting properties of water in
silica nanopores has been studied by previous MD research.\cite{Hart_Rov,
Spohr_Rov, Gal_Chen_2, Gall_SPohr} Such investigations are motivated by the fact
that in experiments the fluid is placed on top of a porous surface and flows
to enter the pores. In the mentioned MD simulations, it was
demonstrated that at all hydration levels water molecules are absorbed by the
Vycor material. However, they have not shown to what extent the pore surface
is wetted, whether it is only partially wetted or rather completely wetted (as
expected from measurements \cite{Grun_Bunt, Vya_Bunt}). Moreover,
the confinement near such hydrophilic surface was found to substantially alter
the dynamic behaviour of water, depending on the filling ratio \cite{Ga_Sp,
Gallo_Spohr}, but it has not been checked how this relates to the
configurations that water can take inside the pore. In all these previous MD
analyses, the (12-6) Lennard-Jones (LJ) potential and the partial charges
assigned to each silica atom site are chosen according to Br{\'o}dka
and Zerda (BZ).\cite{Bro_Zer} In this model, the LJ potential parameters for
silica oxygen atoms are approximated from the Kirkwood-Mueller
formula\cite{Kirk}, while no LJ interaction centers are assigned to silicon
and hydrogen (of the silanol groups) as they are small in size and possess a
low polarizability. In the present investigation, we have found  that water
molecules do not completely wet the BZ silica model surface at intermediate
hydration levels. For this reason, the silica model recently introduced by Gulmen and Thompson \cite{Gulmen_Thompson} (GT) has been tested. The GT
\cite{Gulmen_Thompson} potential is defined similarly to the silica potential
by BZ,\cite{Bro_Zer} however, a weak short-ranged interaction for silicon and
hydrogen atoms has been added and the partial charges on each silica atom are
increased. We probed the performance of both silica models by analyzing the
wetting behaviour of water on silica surfaces and comparing to the experimental results.\cite{Grun_Bunt, Vya_Bunt, Will_Good, Less_Jacob, Los_Jacobs}  \\ \\
In the following, we present the results of MD simulations of water in a
cylindrical silica nanopore of roughly $ 4$\,nm diameter and $ 6.1$\,nm
height. Additionally, MD simulations of water droplets wetting silica slabs of varying thicknesses were performed. The silica nanopores and slabs were
created in our group, with  the silanols molecules uniformly distributed on
the surface. For water in the silica pore, we evaluate the minimum number of
water layers necessary to completely wet a silica surface by looking at the
radial and angular density distribution, as well as the number of hydrogen
bonds formed for different filling ratios of water in the pore. Furthermore, a
phase diagram of the surface tension of the different configurations adopted
by water molecules in the nanopore is calculated analytically, providing deeper
insights into the relation between the interaction energies and the water
arrangement in the pore.\\
 To complete the work, the contact angle of a water droplet on a silica flat
 slab is evaluated and compared between the two model surfaces. Previous MD
 investigations have used the contact angle investigation to approximate LJ
 parameters between water and silica atoms \cite{Cruz_Schu, Emmani_Heinz} but
 they have not stated clearly whether the simulations were performed in such a
 way that the periodic images provided by the periodic boundary conditions do
 not influence the contact angle evaluated. In our investigation, we run the
 simulations without periodic boundary conditions to avoid this issue. 
 Previous experimental results showed that the wetting properties
 do not only involve atoms of layers in the vicinity of the interface but also
 the atoms located deeply inside the slab material. \cite{Will_Good,
 Los_Jacobs, See_Jacobs, Less_Jacob} Therefore, we measure how the contact
 angle changes with the thickness of the slab, showing that a thickness of
$2.5-3$\,nm is sufficient for MD simulations. In order to disentangle the
 contributions of the interfacial electrostatic and van der Waals (VdW) interactions on the
 contact angle, we varied these two contributions in our simulations, showing
 that the influence of the electrostatic interaction is considerably larger
 than that of the VdW interaction.
\section{Simulation details}
 Classical MD simulations were performed with the NAMD \cite{Phil_Sch} 2.10
 simulation package. An amorphous cylindrical nanopore of roughly $ 4 $\,nm of
 diameter and silica slabs of different thicknesses were fabricated in our
 group. To create the pore, a crystalline cell of SiO$_{2}$ with a box length
 of approximately $ 6$\,nm  was built, the system was melted at 5000 K and
 cooled to room temperature with the method described in \cite{Ge_Dro}, and
 then a cylindrical cavity of $ \sim 4 $\,nm diameter was cut.  The process of
 fabrication of the silica cylindrical pore and the silica slab is explained
 in detail in a separate paper.\cite{Ge_Dro} The surface concentration of hydroxyl groups on
 the surface  is $ 7.5$\,nm$^{-2} $ corresponding to highly hydrated silica surfaces. \cite{Bro_Zer} The silica slabs were created following a similar procedure.\\ \\
The silica nanopore and slab contain two types of oxygen atoms depending on
the number of silicon atoms to which they are connected. There are bridging
oxygens (O$_{\mathrm{Si}}$)  bonded to two adjacent silicons and nonbridging
oxygens (O$_{\mathrm{H}}$)  on the  surface attached to only one
silicon. Hydrogen atoms are attached to the O$_{\mathrm{H}}$ in order to form
the silanols groups  (SiOH, Si(OH)$_{2} $). The bonded interaction parameters
for silica atoms were obtained from Hill and Sauer.\cite{Hill_Sauer} Apart from the hydrogen atoms of the silanols groups that are allowed to rotate, all atoms in the silica pore and slab are immobile, constrained to a fixed position. Whereas, water molecules are free to move within the pore.
Liquid water is defined using 2 models: a set of 3 rigid sites given by the SPC/E \cite{Ber_Str} model and a set of 4  sites provided by the TIP4P2005 \cite{Ab_Ve} model. 
The atoms of the silica substrate are allowed to interact with the water sites by means of the Coulomb potential and LJ potential in Eq.~(\ref{eq:one}), 
 \begin{equation}\label{eq:one}
		U_{\mathrm{LJ}} = 4 \epsilon \left( \frac{\sigma^{12}}{r_{i,j}^{12}}  - \frac{\sigma^{6}}{r_{i,j}^{6}} \right)\, 
	\end{equation}
which implements the VdW interaction.
 LJ parameters and fractional charges for the SiO$_{2}$ sites are given in Table~\ref{table:nonlin}.  \\ \\
All simulations were made with the NVT ensemble with a fixed room temperature $ T=298$\, K  using a Langevin thermostat \cite{Phil_Sch} with a coupling coefficient of $ 1.0$\,ps$^{-1} $ and with the hydrogen atoms included in the Langevin dynamics. An integration time step of $ 2 $\,fs was utilized and the simulations were run for at least $ 20$\,ns. Periodic boundary conditions were set for the simulations of water in the nanopore allowing the calculation of the long-range Coulombic electrostatic interactions with the particle-mesh Ewald sum, using a cut-off of $ 1.2 $\,nm. No periodic boundary conditions were defined for the simulation of water wetting silica slabs in order to allow the calculation of the full electrostatic and VdW interactions between all the water droplet atoms and silica slab atoms.
	\begin{table}[ht]
		\caption{LJ potential parameters for silica interaction centers. $^{\star} $values were not mentionned in the model and were chosen arbitrarily small.}
		\centering
		\begin{tabular}{c c c c c}
			\hline\hline
			Parameters & Sites & $\sigma$ (nm) & $\varepsilon$ (Kcal/mol) & q (e) \\ [0.5ex]
			\multirow{4}{4em}{BZ \cite{Bro_Zer}} & Si & 0.0178$^{\star}$ & 0.00000 & 1.283\\ 
			&O$_{\mathrm{Si}}$ & 0.27 & 0.45705694 & -0.629\\ 
			&O$_{\mathrm{H}}$ & 0.3 & 0.45705694 & -0.533\\
			& H & 0.0178$^{\star}$ & 0.00000 & 0.206\\  [1ex]
			\hline
			\multirow{4}{4em}{ GT \cite{Gulmen_Thompson}} & Si & 0.25 & 0.0001 & 1.28\\ 
			& O$_{\mathrm{Si}}$ & 0.27 & 0.457 & -0.64\\ 
			& O$_{\mathrm{H}}$ & 0.307 & 0.17 & -0.74\\
			& H & 0.1295 & 0.0003657 & 0.42 \\ 
			\hline
		\end{tabular}
		\label{table:nonlin}
	\end{table}
\section{Results I: Water in partially filled silica pores}
In the following, we investigated the configuration of water in a partially filled silica pore for the two different models using MD
simulations. Furthermore, we performed an analytical calculation of the
different possible phases of water in a cylindrical pore that allows us to interpret the findings.

\begin{figure}
	\begin{center}
		\includegraphics[width=0.29\textwidth]{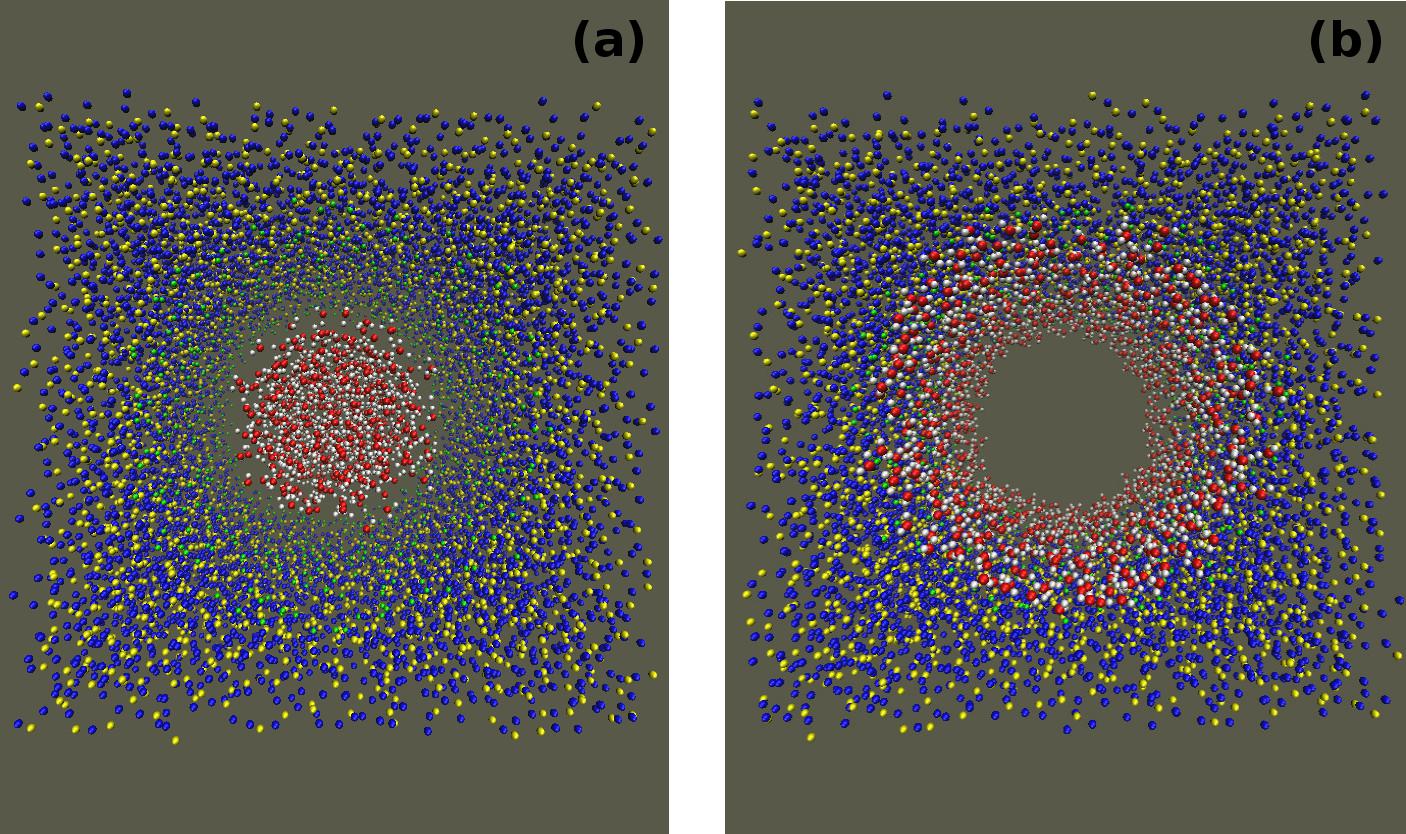}
\end{center}		
	\caption{Top view of the two initial configurations used in the MD
          simulations, labelled in the subsequent figures with "center" (a)
          and "surface" (b). Here the hydration level is $ 30 \% $. Si,
          O$_{\mathrm{Si,H}}$, H are drawn in yellow,  blue and green, while
          water atoms O, H are indicated by red and white respectively. The
          pictures were generated using the VMD program.\cite{Hum_Schul} }
	\label{fig:ini_conf}
\end{figure} 
\subsection{MD simulation results}
\begin{figure}[htbp]
	\begin{center}
			
		\includegraphics[width=0.48\textwidth]{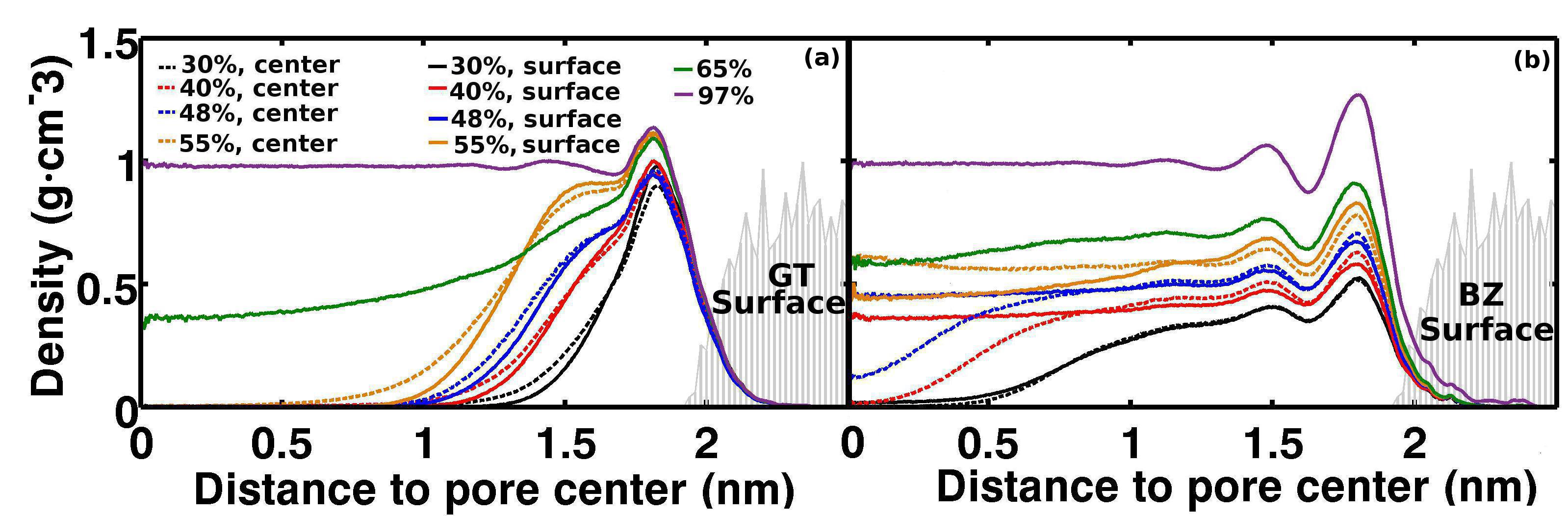}
	\end{center}
	\caption{Radial density profile of water in the silica pore for
          different filling ratios and starting configurations, for the (a) BZ surface and the (b) GT surface. The radius of the pore is 2\,nm. The gray area indicates the silica pore surface and is arbitrarily scaled for a clear visibility.}
	\label{fig:raddens_cs}
\end{figure}
\begin{figure}[htbp]
	\begin{center}
			\includegraphics[width=0.48\textwidth]{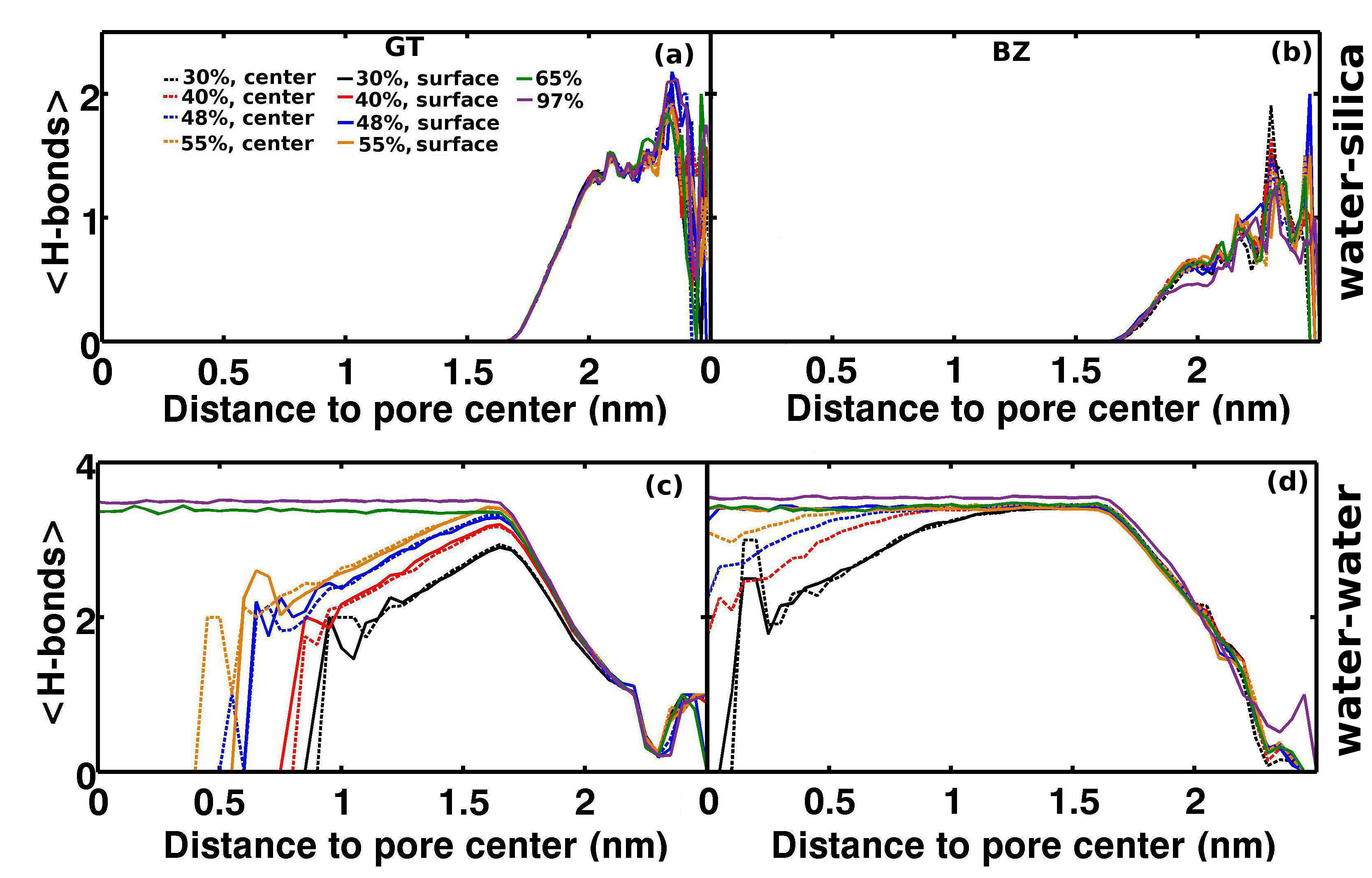}
	\end{center}	
	  	\caption{Average number of hydrogen bonds per water molecules
                  for water--silica (a-b) and water--water (c-d) contacts, for
                  different initial conditions and filling ratios, as
                  indicated in the legends. The results for the GT silica
                  surface are given on the left-hand side, and those for the
                  BZ silica surface on the right-hand side.}
	  	\label{fig:diff_conf}
	  \end{figure}

We evaluated the equilibrated configurations of water in the pore for the GT and BZ surfaces, using different filling ratios and different starting configurations. The pore filling ratios are in the range  30$\%$ -  97$\% $, based on the estimated number of molecules for 100$\% $ filling ratio, which is 2700.\cite{Spohr_Rov}
The equilibrium configurations were analyzed by calculating the radial density profile, the
distribution of water molecules on the interior pore surface, and the number of hydrogen bonds among water molecules and between water and silica molecules. In order to see how far the final configurations depend on the initial configuration, we used the two different initial configurations shown in Fig.~\ref{fig:ini_conf}, where water is concentrated around the cylinder axis and at the pore surface, respectively. There is thus a void between the water droplet and the silica surface in the first configuration, and a void in the pore center for the second configuration. \\ 
Fig.~\ref{fig:raddens_cs} shows the radial density profile of water molecules inside the pore as function of the distance to the pore center, averaged over 15\,ns after at least 5\,ns of equilibration for each simulation.
One can see that water is closer to the GT surface.
The GT density profile shows only one peak for a filling ratio of 30$\%$,
indicating that all water molecules are in contact with the pore surface. Only
after the first layer is completed, a second layer is formed, as is visible
for the curves for filling ratios  between 40 and 55$\%$. At 65$\%$ filling ratio, the water molecules can also
be found in the interior of the pore, indicating a configuration with
a completely wetted surface and a compact water droplet in the pore interior. 
\begin{figure}[htbp]
	\begin{center}
			\includegraphics[width=0.48\textwidth]{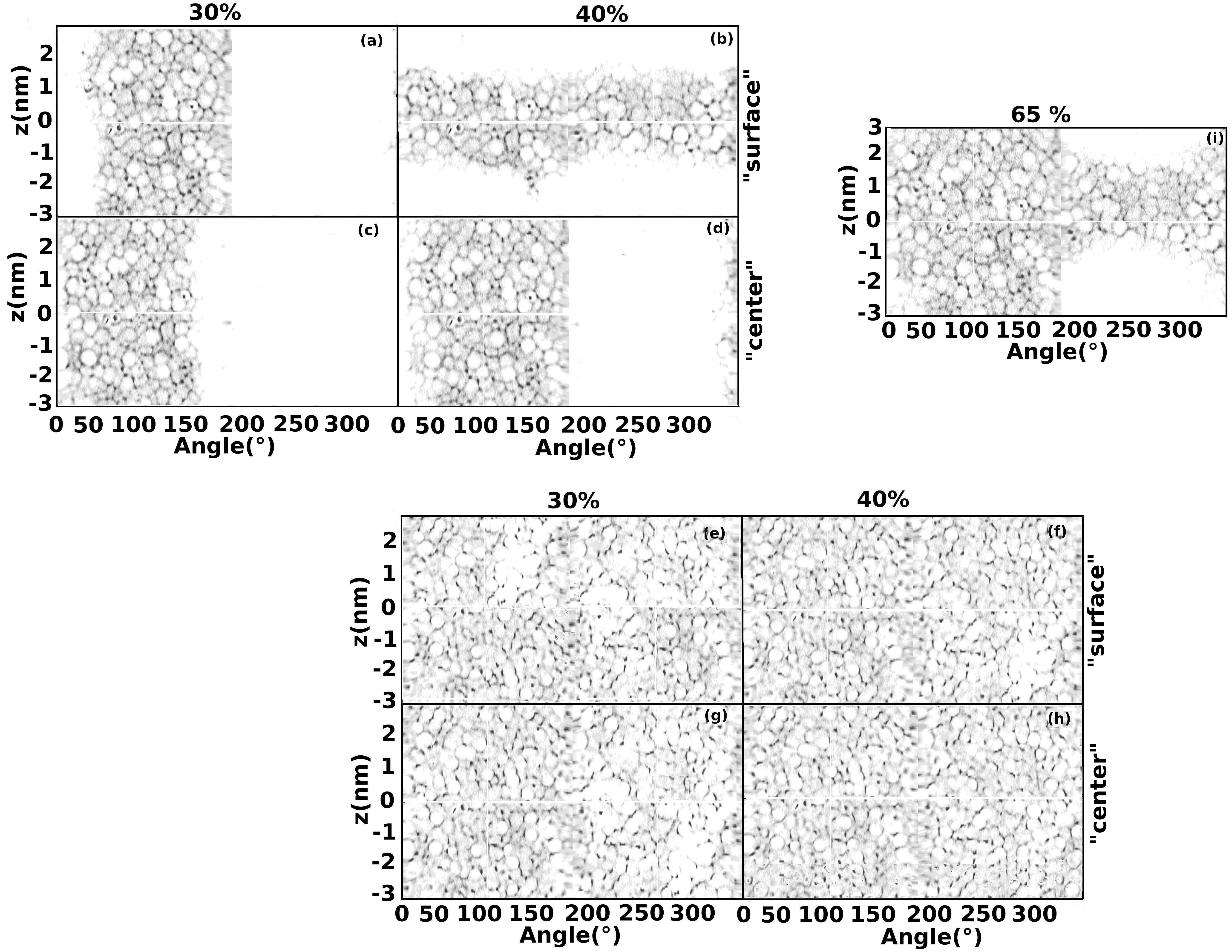} 	
	\end{center}
	\caption{The water density at the pore surface, showing water
          molecules that are within a distance of 0.3\,nm of the silica surface. For the GT surface (e-h), only filling ratios of 30 and 40$\%$ are shown, since for larger filling ratios the surface is completely wetted. For the BZ surface, the top (a-b) graphs correspond to the "surface" starting configuration and the bottom graphs (c-d) correspond to the "center" starting configuration, (i) represents the density at 65$\%$ filling ratio.}
	\label{fig:surfdens_cs}
\end{figure}
The density profiles for the BZ surface show several layers of water, with a
peak height that depends on the filling ratio. Furthermore, the density
profile depends on the initial
configuration for intermediate filling ratios, with the initial configurations at
the surface leading to final configurations with rather flat density
profiles. This suggests that for the initial configuration at the boundary,
the water droplet forms a "plug" in the pore interior, while for the initial
configuration in the center, water forms some type of droplet sitting at
the surface. Since the pore surface is rough, some water molecules can also be found inside the silica pore material. \\ 
In order to test the intuition obtained for the water configurations based on the density profiles, we evaluated the distribution of water molecules within a distance of 0.3\,nm of the surface. Fig.~\ref{fig:surfdens_cs} shows the
resulting surface density profiles, using cylinder coordinates.
This figure confirms that for the GT surface, the water droplet first wets the
surface completely, before filling the interior. For the BZ surface, the pore
surface is only partially covered with water, and the final configuration
depends on the starting configuration for intermediate filling ratios. 
For instance, for 40$\%$ filling ratio with the "center" starting configuration,
water molecules are concentrated in one angular segment of the surface but are
covering the whole length, while for the "surface" starting configuration
they occupy only part of the $z$ range, but all angles. These final configurations for intermediate filling ratios are in fact very plausible if one tries to imagine how the initial configurations can evolve with time in a situation where the water-surface interaction is not strong enough that the entire surface is wetted. When the initial
configuration has a water cylinder in the pore center, the entire cylinder gets attracted by the silica molecules under the influence
of electrostatic and VdW interaction and moves as a whole towards the pore
surface, wetting a specific angular region of the surface. 
When the initial configuration sits at the pore surface, the water film may
rupture along an angular line, and the water will contract to form a
plug. Even if one of the two final configurations has a lower free energy,
this free energy difference will not be large, and the transition between them
will involve a barrier that is so large that it is not overcome during the
simulation time.
When the filling
ratio is lower (as can be seen for 30$\%$), the plug is not observed for either
initial configuration, indicating that there is only one stable configuration.
 The two different final configurations merge also for larger filling ratios (as
 can be seen for 65$\%$), where the void left by the water droplet takes the shape of a droplet that sits at the pore surface.\\
Finally, we evaluated the average number of hydrogen bonds formed between
water molecules, and between water molecules and silica molecules. This shows
to what extent the stronger hydroaffinity of the GT model affects the
formation of molecular bonds. We considered two oxygen atoms to be connected via a hydrogen bond if the angle between the intramolecular O-H vector and the intermolecular O...O  vector is less than $30^{\circ}$, provided that the  O...O  separation is less than $0.335$\,nm. The results are shown in Fig.~\ref{fig:diff_conf}.
For the BZ surface, the number of hydrogen bonds between water molecules reaches the bulk value in the inner part of the pore for a filling ratio larger than 40$\%$ with the "surface" initial configuration. Also for the
"center" initial condition, the bulk value is reached for the  water molecules in the interior of the water droplet. In the GT surface, the bulk  value is reached only for filling ratios above 60$\%$. This illustrates the fact that the BZ surface disrupts the water structure more than the GT
surface. Accordingly, the number of hydrogen bonds formed between the silica surface and the water molecules is larger for the GT surface. \\
For both models, the maximum number of water-silica hydrogen bonds is already reached at  $40 \%$ filling ratio, confirming that one and half layer of water molecules is sufficient to completely wet the GT silica surface. It is at first surprising that for the BZ surface, the number of water-silica hydrogen bonds does not increase for filling ratios larger  $40 \%$ and stays considerably below the value of the GT surface. This can only be explained by different water orientations near the surface in the two models. In the supplementary
information 
we show that near the BZ surface the OH bonds of water molecules have a
preferred orientation, while this is not the case near the GT surface. This
means that only part of the  water molecules can act as a hydrogen bond donor or acceptor near a BZ silica molecule, while near the GT
surface all the water molecules can share
hydrogen bonds with the surface atoms.

\subsection{Theoretical evaluation of the phase diagram}\label{phases}
In order to better understand the dependence of the water droplet configuration in the pore on the interaction energies and the filling ratio, we performed a theoretical analysis that is based on surface energy minimization. Denoting the surface area between the water droplet and vacuum with $A_{\mathrm{1}}$, the surface area between the water droplet and the pore material with $A_{\mathrm{2}}$, and $\gamma_{\mathrm{1}}$ as the surface tension between water and vaccuum, $\gamma_{\mathrm{2}}$ as the difference between the surface tension of silica and water  with the surface tension of silica and vacuum. The total surface energy of the wetting droplet can be written as
\begin{equation} \label{eq:surfaceEnergy}
   E_{\mathrm{S}}= \gamma_{\mathrm{1}} \cdot A_{\mathrm{1}}+ \gamma_{\mathrm{2}} \cdot A_{\mathrm{2}}\, .
\end{equation}
If we assume that the entropy does not change much between different phases, the configuration of the water droplet in the pore can be obtained by minimizing $E_{\mathrm{S}}$ for a given filling ratio. In order to perform the calculation mostly analytically, we approximated the different possible phases using simple geometrical shapes, so that the energy minimization can be performed by varying at most 2 parameters that characterize the phase. We fixed the  ratio of the pore radius and pore length to the value $R/L=2/6.1$ used in the simulations. We determined the phase diagram in dependence of the filling ratio and the ratio between the two surface energies. We allowed for hydrophilic ($\gamma_{\mathrm{2}}<0$) as well as for hydrophobic surfaces ($\gamma_{\mathrm{2}}>0$). The surface tension $\gamma_{\mathrm{1}}$ is a positive quantity. 

\begin{figure}[htbp]
\begin{center}
    \includegraphics[width=0.48\textwidth]{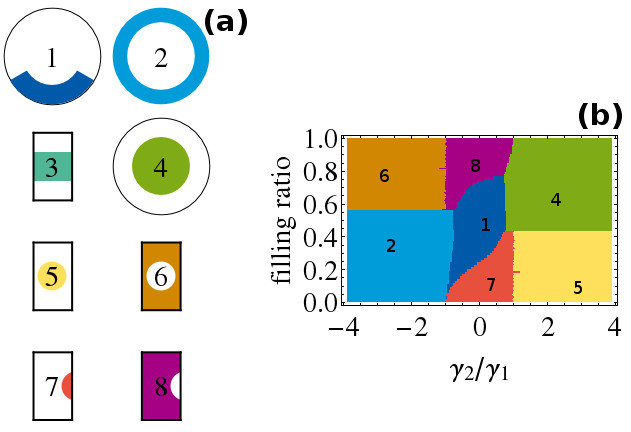}
  \end{center}
    	\caption{(a) The eight different phases used for energy minimization and (b) the phase diagram in dependence of the ratio of the water-vacuum and water-silica surface energies and of the filling ratio. The aspect ratio between radius and length of the cylindrical pore is 2/6.1.}
  	\label{fig:phaseDiagram}
\end{figure}

Fig.~\ref{fig:phaseDiagram} shows the eight phases and the phase diagram obtained from  minimizing $E_{\mathrm{S}}$. 

Phases 1,2, and 4 are translationally invariant along the cylinder axis and represent the cases of partial, full, and no wetting of the silica surface. Phase 3 represents a plug in the shape of a cylinder that is shorter than the pore. Phases 5-8 describe the cases where the water or the vacuum forms a spherical droplet in the interior or one that intersects with the pore surface. 

The calculation of the surface energy for phases 1-6 is a straightforward
analytical calculation. For phase 1, we had to use mathematica to evaluate the
final expression. In order to evaluate phases 7 and 8, we had to resort to a
numerical evaluation. We first created a data base by calculating numerically
the volume and surface of the cut droplet for over 1 million different combinations of the sphere radius and the distance of the sphere center from the cylinder axis. Then, we used this data base to find for a given filling ratio the values that minimize $E_{\mathrm{S}}$. 

The phase diagram shows clearly three qualitatively different regions that
depend on the ratio $\gamma_{\mathrm{2}}/\gamma_{\mathrm{1}}$. For $\gamma_{\mathrm{2}}/\gamma_{\mathrm{1}} \lesssim -1$, the energy is minimized by having maximum surface area with the pore surface. The water wets the pore completely. Correspondingly, the phases 2 and 6 occur. With increasing filling ratio, the volume of the vacuum becomes smaller, and eventually a free volume that does not touch the surfaces fits into the pore. For sufficiently large filling ratio, a droplet clearly has the smaller surface area with the vacuum and therefore the lower energy. (Our simple calculation did not take into account that the droplet could be stretched, and therefore the position of the phase boundary between phases 2 and 6 should in fact be at a lower filling ratio.)

For $\gamma_{\mathrm{2}}/\gamma_{\mathrm{1}} \gtrsim 1$, the pore material is highly hydrophobic, and the phases 4 and 5, which have no contact between water and silica, have the lowest energy, depending on the filling ratio. 

In the intermediate parameter region $-1\lesssim\gamma_{\mathrm{2}}/\gamma_{\mathrm{1}}\lesssim 1$, we observe phases 7, 1, and 8 as the filling ratio is increased. These are the phases that have surfaces with the pore and with the vacuum. Since the (absolute value of) water-vacuum energy is larger than that of the water-silica energy, these phases are to a large extent affected by the condition that the water-vacuum interface shall be minimum. The transition from phase 7 to phase 1 occurs for lower values of $\gamma_{\mathrm{2}}/\gamma_{\mathrm{1}}$ at smaller filling ratios than for larger $\gamma_{\mathrm{2}}/\gamma_{\mathrm{1}}$, because a larger surface area to the pore is energetically favorable for negative $\gamma_{\mathrm{2}}$. For the same reason, phase 8 wins over phase 1 for high filling ratios and negative $\gamma_{\mathrm{2}}$, because phase 8 has more surface area between water and the pore.

Phase 3 does not occur in the phase diagram. It will certainly occur when the
ratio between radius and length of the pore becomes smaller, because it has
then  smaller surface areas than phase 1. In our simulations with the BZ
potential, we found this phase for intermediate filling ratios, where it
coexists with phase 1. Phase 3 thus  might well be metastable. On the other
hand, it is also possible that phase 3 is indeed stable in part of the phase
diagram due to entropic effects, which were not taken  into account when
calculating the phase diagram. 
Since in the canonical NVT ensemble the free energy $F=E-TS$ has to be
minimized, phases with larger entropy become more favored when entropy is
taken into account. This will shift the phase boundaries somewhat. For
instance, when phase 2 contains only two layers of water molecules, its
entropy per molecule is smaller than in bulk water. Similarly, the entropy per
water molecule is larger in phase 3 than in phase 1, since the water in phase
3 is more bulk-like. 

With the insights gained from these analytical calculations, we can interpret
the results of the MD simulations: For the BZ surface, the water wetted the
silica surface only partially for all simulated filling ratios, and we
observed the phases 1, 3, and 8 depending on the filling ratio. This means that the ratio of surface energies in the interval $\gamma_{\mathrm{2}} / \gamma_{\mathrm{1}} \in (-1,0)$. (Since the surface is hydrophilic, we have $\gamma_{\mathrm{2}}<0$.)  
 With the GT surface, we observed a complete wetting of the silica surface
 (phases 2 and 6) for all simulated filling ratio, which means that $\gamma_{\mathrm{2}} / \gamma_{\mathrm{1}} < -1  $. This appears to be the more realistic scenario, as it agrees well with experimental results.\cite{Grun_Bunt} 

In order to obtain an additional perspective on the different interaction
between water and a silica surface in the two models, we will in the next
section investigate the contact angle of water on top of a flat silica slab using both models.  
\section{Results II: Water on top of a silica slab}
\begin{figure}[htbp]
 \begin{center}
\includegraphics[width=0.48\textwidth]{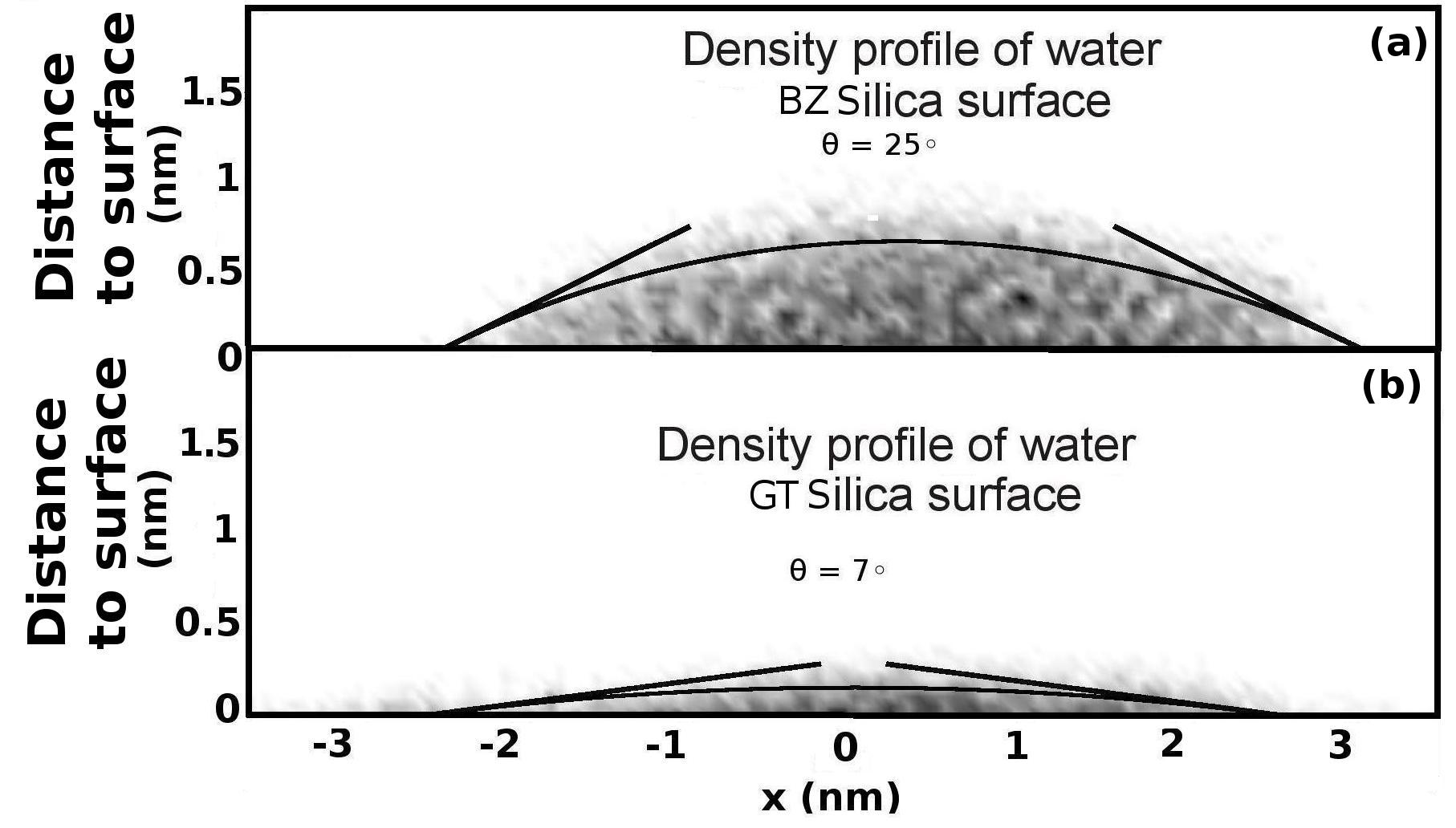}
  \end{center}
 	\caption{Contact angle of a water droplet on top of a silica slab of thickness $t=2.5 $\,nm. (a) and (b) correspond respectively to the BZ and the GT silica surface.}
  	\label{fig:ca_GB}
\end{figure}

A good tool to examine the performance of silica potentials is the evaluation
of the contact angle of a water droplet wetting the surface. We performed MD
simulations of a water droplet on a flat surface of amorphous silica and measured the
contact angle. We did not use periodic boundary conditions in order to remove the influence of the neighbouring water periodic images. Instead, the Coulombic and the VdW (LJ) interaction energies between all atoms in the water droplet and the silica slab were calculated exactly.\\
In order to evaluate the contact angle $ \theta$, the density profile of all horizontal water layers of $0.05$\,nm thickness was determined, and from these a contour plot of the density was obtained. The contour plot was fitted to a circular segment, and the contact angle was deduced from the tangential line to the base of the circular segment. The result is shown in
Fig.~\ref{fig:ca_GB} for both types of potentials, with a slab of thickness $t= 2.5$\,nm. For the GT model, the droplet covers the entire surface and has a very small contact angle of $7^{\circ}$. When we performed the same simulation with periodic boundary conditions, the water layer became completely flat. In contrast, the contact angle of the water droplet on top of the BZ surface is $25^{\circ}$. These results confirm the findings of the previous subsection, that the GT silica surface is so hydrophilic that
water wets it completely, while the BZ surface is less hydrophilic. 

The contact angle is closely related to the surface tensions that we used for
evaluating the phase diagram. The condition that the total surface
energy ($ E_{\mathrm{S}}$) of water wetting a silica surface must be minimal
for an equilibrated droplet of constant volume, i.e., $\mathrm{d}E_s=0$, leads to 
\begin{eqnarray}\label{eq:Es}
		\frac{\mathrm{d}A_{1}}{\mathrm{d}A_{2}}=
                -\frac{\gamma_{2}}{\gamma_{1}} = \cos \theta \, .
	\end{eqnarray}
The contact angle of $25^{\circ}$ for the BZ surface agrees with our conclusion
that $-1 <\gamma_{2}/\gamma_{1}<0 $ and leads more precisely to
$\gamma_{2}/\gamma_{1}\simeq -0.9$.
For the GT surface, we had obtained that $\gamma_{\mathrm{2}} /
\gamma_{\mathrm{1}}< -1 $, implying a contact angle of zero and complete
wetting, which is compatible with our finding that the droplet spreads on the
surface forming only the 1.5 layers required for wetting.

Our finding of a very small contact angle for the BZ surface is in agreement
with what is found in experiments.\cite{Will_Good,Less_Jacob, Los_Jacobs} In
those experiments, the precise value of the angle depends on whether the
advancing or receding droplet is considered, and on the thickness of the
slab, and it ranges from  $0^{\circ}$ to  $7^{\circ}$. 

\begin{figure}[htbp]
\begin{center}
	\includegraphics[width=0.48\textwidth]{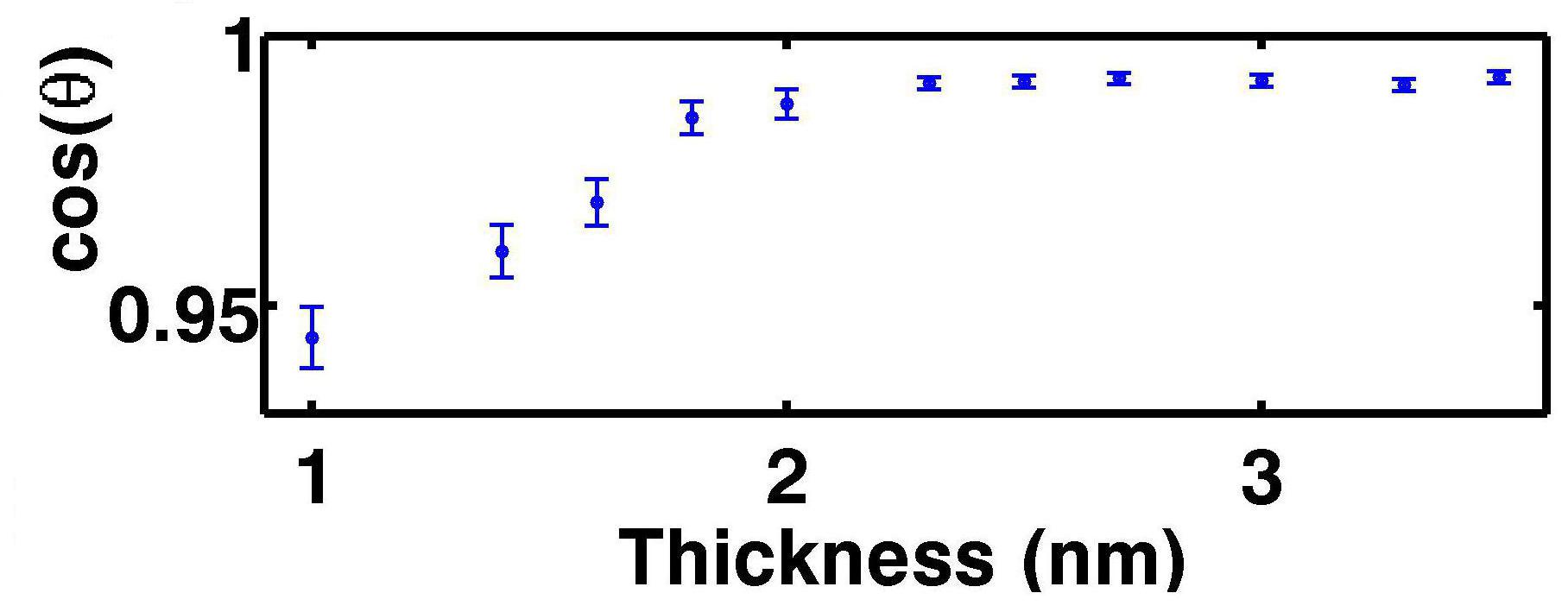}
\end{center}
 	\caption{Dependence of the cosine of the contact angle $\theta$ of the water droplet on the thickness of the silica slab. The MD simulation was done with the GT model, using 469 water molecules.}
 	\label{fig:ca_sim}
 \end{figure}

Since the interaction strength between water and silica is known to depend on the thickness of the slab \cite{Will_Good, Los_Jacobs, See_Jacobs, Less_Jacob}, we evaluated the contact angle for different values of the thickness $t$ (within the GT model), see Fig.~\ref{fig:ca_sim}. The cosine of the contact angle increases with thickness and reaches its limit
when $t=2.5$\,nm, where the contact angle is $\approx 7^{\circ}$. A similar
dependence of the cosine of the contact angle on thickness has been measured.\cite{Will_Good} It means that, to accurately reproduce the wetting
properties of a water droplet on a silica slab, the slab should have at least
a thickness in the range 2.5-3\,nm. This result is close to the experimental
value 3\,nm found by Williams and Goodman.\cite{Will_Good}
The change of the contact angle with thickness shows also that water does
not wet the surface completely when the slab is thin. This means that the
ratio $\gamma_{\mathrm{2}} /\gamma_{\mathrm{1}} $ changes from a value $>-1$
to a value $<-1$ with increasing thickness, implying that this ratio is very
close to -1. This in turn means that small differences in the preparation of
the surface can shift the ratio over the threshold -1, and this may explain why some experiments find a nonzero contact
angle, while others find complete wetting.
 \begin {table}[H]
\begin{center}
	\begin{tabular}{ | l | c | c | c |}
		\hline
		\small & \small $ \epsilon_{\mathrm{GT}}, q_{\mathrm{GT}}  $ & \small $1 \%  \epsilon_{\mathrm{GT}}, q_{\mathrm{GT}}  $  & \small $ \epsilon_{\mathrm{G}}, (q=0 ) $   \\ \hline  
		\small	  $t=1.6$\,nm  & $14.3^{\circ}$  &  $17.2^{\circ}$ & $105^{\circ}$\\ \hline
		\small	 $t=2.0$\,nm   & $9.0^{\circ}$ &  $12.3^{\circ}$  & $88^{\circ}$\\ \hline
		\small	 $t=2.5$\,nm  & $7.0^{\circ}$ & $11.1^{\circ}$ & $87^{\circ}$ \\ \hline
		\hline
	\end{tabular}
	\caption {Contact angle obtained for different slabs types:  $  \epsilon_{\mathrm{G}T}, q_{\mathrm{GT}}$ represent the GT model surface as defined in table~\ref{table:nonlin}. $1 \%  \epsilon_{\mathrm{GT}}, q_{\mathrm{GT}}  $  represents the GT model in which $\epsilon $ of each atoms is reduced to $ 1 \%  $ of the original value in the GT model; $ \epsilon_{\mathrm{GT}}, (q=0 ) $ stands for a slab without partial charges assigned to atoms but with LJ sites corresponding to the GT model }
	 \label{tab:CA}  
\end{center}
\end {table}

In the following, we evaluate the influence of the different types of
interaction on the contact angle, since previous investigations
\cite{Los_Jacobs, Less_Jacob} have shown that the VdW interaction plays an
important role in the adhesion of biomolecules on surfaces, in particular silica surfaces.
To this purpose, we evaluated the contact angle for the case where the partial
charges of silica atoms were set to 0 (this means no electrostatic interaction between water and silica), and for the case where the LJ interaction
was
reduced to 1 percent of its original strength. Table~\ref{tab:CA} compares the
resulting contact angles for different slab thickness. 

It is interesting to see that without partial charges, one obtains a contact
angle of $105^{\circ}$ for thin slabs, which means that the surface has become
hydrophobic, with $\gamma_{\mathrm{2}} /\gamma_{\mathrm{1}} \simeq 0.26 $. But
it becomes slightly hydrophilic when the slab is thicker, and this change in the hydroaffinity must be due to the VdW interaction.

In the opposite case, where the VdW interaction is reduced 
the surface is highly hydrophilic ($ \theta = 11.1^{\circ}$), even though it
is not as hydrophilic as with the VdW interaction. This means that
the electrostatic interaction influences the wetting properties more than the
VdW interaction.

\section{Conclusion}
We have investigated the wetting properties of water in partially filled
silica pores and on flat silica surfaces using MD simulations with two widely
used  intermolecular potentials \cite{Gulmen_Thompson, Bro_Zer} for the
interaction of water and silica surface atoms. This investigation was made
possible since we have developed a method for contructing realistic silica
slabs \cite{Ge_Dro}, with uniformly distributed silanols group on the
surface. The study was motivated by experimental findings that water in
partially filled silica pores wets the surface
completely.\cite{Grun_Bunt,Vya_Bunt} These experiments measured the absorption
of water in MCM-41, SBA-15 \cite{Grun_Bunt} and in CPG \cite{Vya_Bunt} pore
material using $ ^{1}\mathrm{H} $-MAS solid state NMR spectroscopy. 
An evaluation of the $ ^{1}\mathrm{H}$ chemicals shift of the water
hydrogen atoms in
the pore showed that the inner surface of the pore is at first completely
covered during the filling mechanism, and with increasing filling ratio the
center of the pore becomes filled. These experiments were used to measure the surface density of silanol groups on the pore surface. 

The finding that water wets the pore surface completely was not retrieved in our first simulations using 
the standard BZ silica model.\\ Our comprehensive analysis shows that the GT silica model
\cite{Gulmen_Thompson} surface is more hydrophilic than the standard BZ silica
model \cite{Bro_Zer} surface. This must be due to the different partial
charges assigned to silica atoms in the two models, in agreement with our result
shown in the last section that the electrostatic interaction plays a prominent
role in the wetting properties of water on silica surfaces. \\ Based on the
evaluation of radial and angular density profiles,  we found that the GT
silica model surface is completely covered by water molecules even at low
filling ratios (30 $\%$, 40 $\%$), and that roughly one and a half layer of water is sufficient for a full wetting of the surface.
This result does not show a dependence on the water models used and on the starting configuration of the water droplet inside the silica pore. 
For the BZ model surface, the arrangement of water in the pore depends on the
filling ratio and on the starting configuration. A simple analytical
evaluation of the possible water phases in a pore revealed that for the BZ
model, the interfacial energy of the water droplet with the silica wall
is smaller than the interfacial energy of water
with vacuum, leading to water configurations that have small interfaces with
the vacuum, while the opposite is true for the GT surface, where the interface
with the silica surface is maximized. 
\\ 
Near the BZ surface, we observed a formation of a double water layer at all
filling ratios studied, agreeing well with the findings of other authors.\cite
{Hart_Rov, Spohr_Rov, Gall_SPohr} However, our findings complement their work
as we evaluated also the proportion of the pore surface that was wetted, and
demonstrated the coexistence of different phases of water inside the pore at
intermediate filling ratios. The occurring phases could not be fully explained
by the phase diagram that was based solely on surface tensions, and we argued
that the effect of entropy makes it plausible that the more compact "plug"
phase is favored in the MD simulations. This conclusion finds support from MD
simulation results \cite{Gal_Chen_2,
  Gallo_Spohr, Gall_SPohr, Ga_Sp} (using again the BZ model) that for filling ratios $ \leq $ 56 $\%$ water molecules within the
first two layers from the substrate are in a glassy state even at room
temperature, which means that they contribute little to the entropy.
 \\
In agreement with the different water arrangements found for the two
potentials, we found that the contact angle of a water
droplet wetting a silica slab is close to zero for the GT silica model, while
it is around 25$^{\circ}$ for the BZ model. Again, the GT model agrees well with experimental results.\cite{Less_Jacob, Los_Jacobs}  Furthermore, the contact angle decreases with slab thickness and reaches its limit at approximately $t=2.5$\,nm. This result matches well with the  experimentally found value $t=3$\,nm \cite{Will_Good} and can
help in decreasing the computational cost of MD studies by limiting the
thickness of the slab to $t=3$\,nm. Our simulations show also that the surface
energy ratio $\gamma_{\mathrm{2}} /\gamma_{\mathrm{1}} $ is very close to -1
and can shift from a value $>-1$ (implying a finite contact angle) to a value
$<-1$ (implying zero contact angle) with increasing thickness. Therefore small
changes in  the preparation of the silica surface can move the ratio over the
threshold -1 and may explain why  a nonzero contact angle is found in some experiments
\cite{Less_Jacob, Los_Jacobs} and complete wetting in others.\cite{Will_Good,Grun_Bunt,Vya_Bunt} 
  \\
Finally,  varying the LJ parameters and partial charges on the silica
atoms, we found that turing on the VdW interaction 
 can make a hydrophobic non-polar surface slightly hydrophilic, underlining
 the important role of the VdW interaction found in experiments.\cite{Los_Jacobs, Less_Jacob}  However, electrostatic interactions play a dominant role for the wetting properties of water on polar surfaces.

\begin{acknowledgments}
This work was supported by DFG
grant number Dr300/11-2. It was also supported in part by Perimeter
Institute for Theoretical Physics. Research at Perimeter Institute is
supported by the Government of Canada through Industry Canada and by the
Province of Ontario through the Ministry of Economic Development \& Innovation.
\end{acknowledgments}


\end{document}


\preprint{AIP/123-QED}

\title{On the influence of the intermolecular potential on the wetting properties of water on silica surfaces\footnote{Error!}}
\author{E. Pafong}
\affiliation{Institut f\"ur Festk\"orperphysik, Technische Universit\"at 
Darmstadt, Hochschulstr. 6, 64289 Darmstadt, Germany}
\thanks{priscel@fkp.tu-darmstadt.de}
\author{J. Geske}
\affiliation{Institut f\"ur Festk\"orperphysik, Technische Universit\"at 
Darmstadt, Hochschulstr. 6, 64289 Darmstadt, Germany}
 \author{B. Drossel}
\affiliation{Institut f\"ur Festk\"orperphysik, Technische Universit\"at 
Darmstadt, Hochschulstr. 6, 64289 Darmstadt, Germany}
\date{\today}
\maketitle
\section{Water in 97 $\% $ filled silica pore}
\begin{figure}[htbp]
\begin{center}
\includegraphics[width=0.5\textwidth]{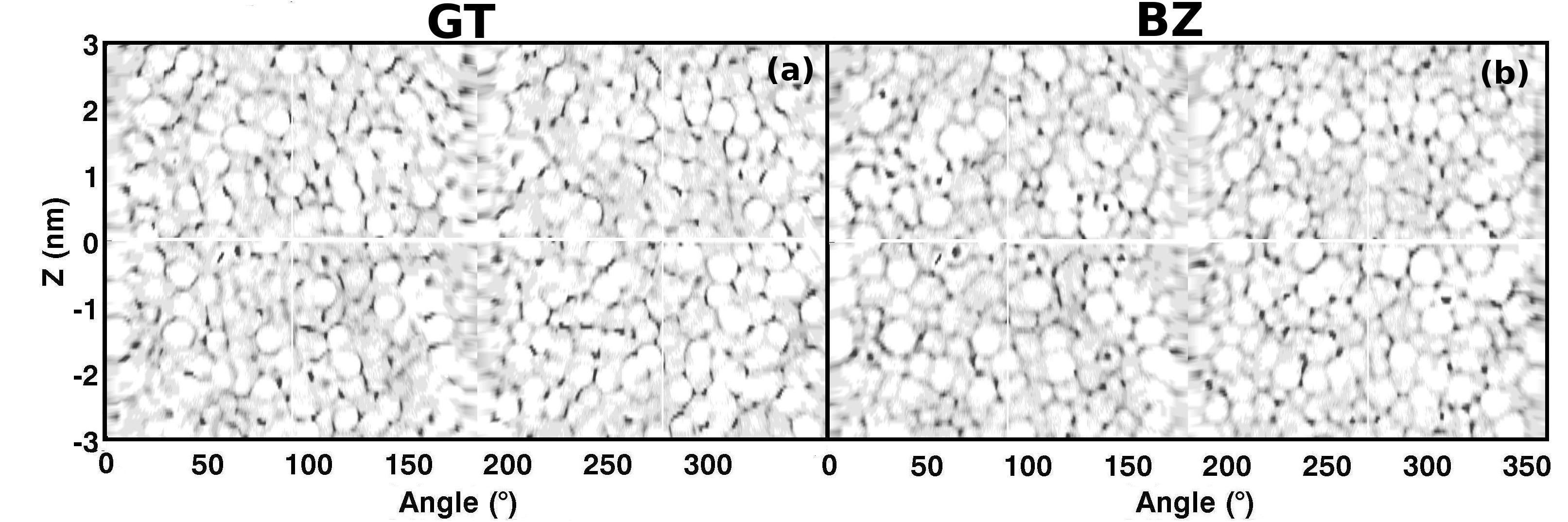}
\end{center}
\caption{The water density at the pore surface for 97 $\% $ filling ratio, showing water molecules that are within a distance of 0.3\,nm of the silica surface, (a) shows the density on the GT surface and (b) on the BZ surface.}
	\label{97_BG}
	\end{figure}
\begin{figure}[htbp]
\begin{center}
\includegraphics[width=0.5\textwidth]{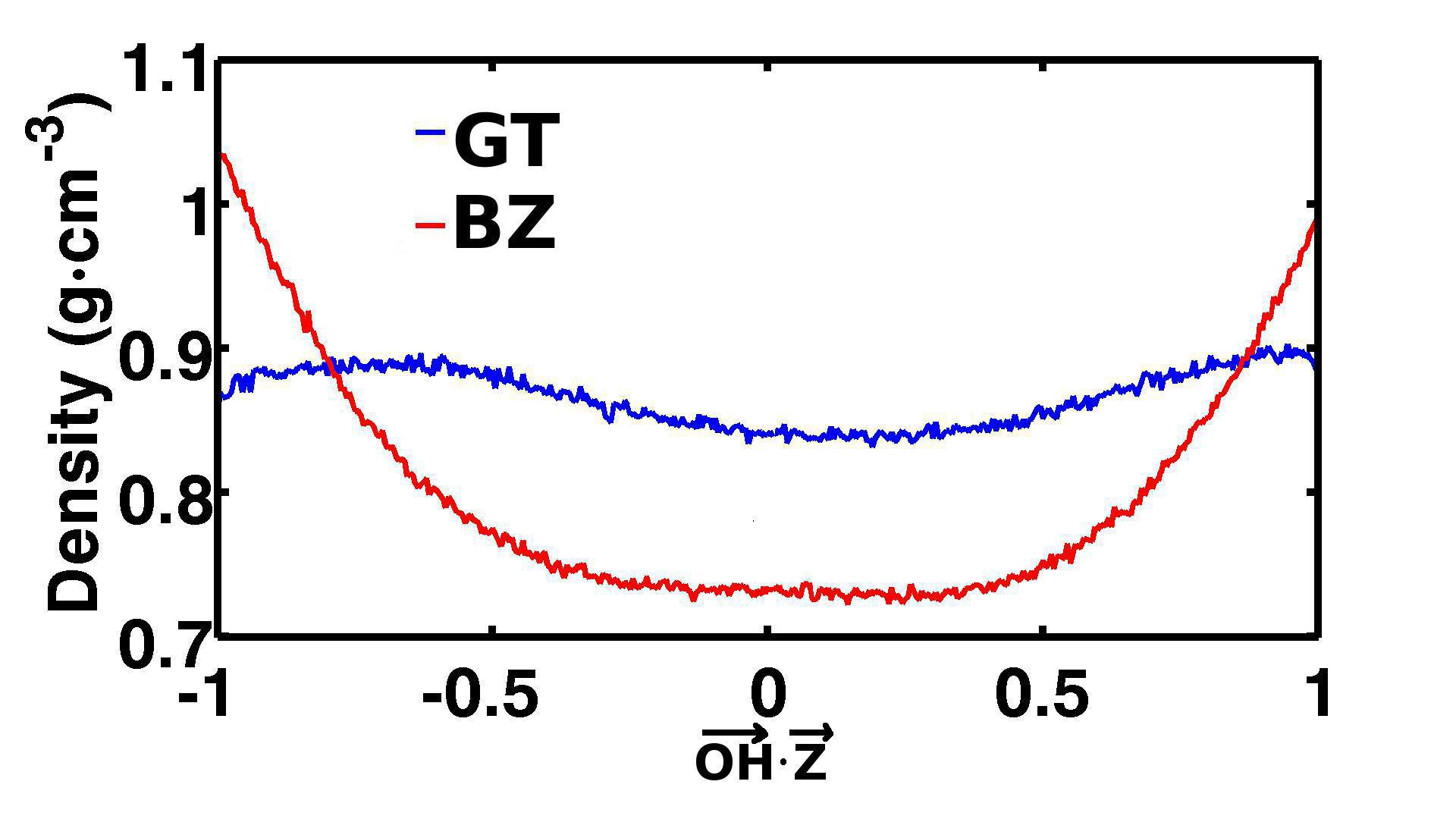}
\end{center}
\caption{Density distribution of the scalar product between a $ \overrightarrow{\mathrm{OH}}$ vector of the water molecules and a unit vector of the z-axis.}\label{angz}
	\end{figure}

The distribution of water molecules within a distance of 0.3\,nm of the pore surface, using cylinder coordinates, is plotted for each model surface in Fig.~\ref{97_BG}. We can notice that in both model surface, the surface is completely covered, tough it has been shown in the paper that there are always less hydrogen bonds formed between water and silica atoms of the BZ surface.
An investigation of the orientation of the $ \overrightarrow{\mathrm{OH}}$ vector of the water molecules (lying within $ 0.3 $\, nm of the pore surface) with respect to the pore axis (z-axis) has been performed. The density profile of the scalar product $ \overrightarrow{\mathrm{OH}}\cdot\vec{\mathrm{Z}}$ has been plotted in Fig.~\ref{angz}.
In the BZ model surface, the $ \overrightarrow{\mathrm{OH}} $ vectors prefer to orient themselves parallel to the z-axis and this preference reduces the number of hydrogen bond donors which can form hydrogen bonds with the oxygen acceptors lying in the silica surface. This preference is not observed near the GT model surface, there is almost no preferential orientation and thus all water molecules have an equal probability to act as hydrogen bond donors or acceptors for the silica hydrogen bond sites. 

\section{Influence of the water model}
In order to compare the performance of both silica potential models for different water models, we have chosen the SPC/E water model \cite{Ber_Str}  which is a three interaction sites model and the TIP4P2005 \cite{Ab_Ve} which is a four sites water model. We have found that the average number of hydrogen bonds shared between water molecules for a TIP4P2005 water droplet is 3.6 whereas it is 3.5 for a SPC/E water droplet. Thus, the TIP4P2005 water model provides water molecules which are slightly more polarized than the SPC/E water molecules. In order to investigate whether this contrast will create a difference in the amount of hydrogen bonds shared with the silica substrate, MD runs have been performed for TIP4P2005 and SPC/E water confined in the cylindrical pore for different filling ratios. 
\begin{figure}[htbp]
\begin{center}
\includegraphics[width=0.45\textwidth]{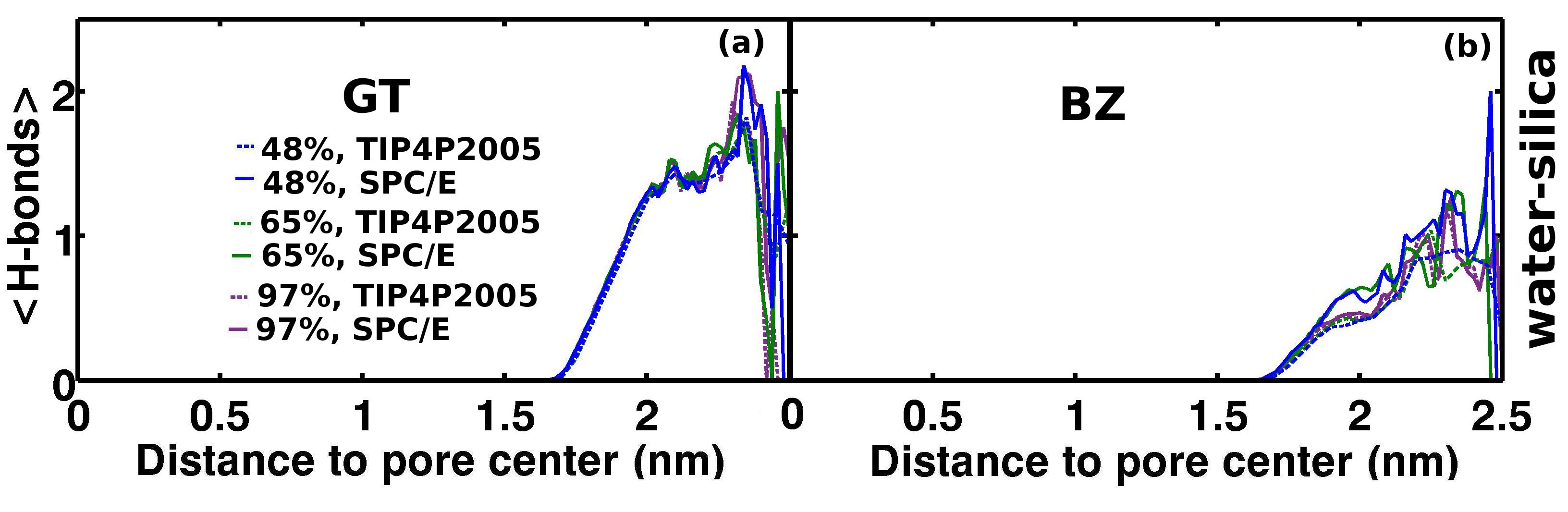}
  \end{center}
    	\caption{Average number of hydrogen bonds per water molecules
                  for water--silica contacts, for
                  different filling ratios and water models, as
                  indicated in the legends. The results for the GT silica
                  surface are given on the left-hand side (a), and those for the BZ silica surface on the right-hand side (b).}
  	\label{diff_wmodel}
\end{figure}
It can be seen in Fig.~\ref{diff_wmodel} that either close to the GT or the BZ model surface, the number of hydrogen bonds formed between water molecules and atoms of the silica surface show a similar trend for both water models. However, there are a little bit more water--silica hydrogen bonds shared with the SPC/E water molecules than with the TIP4P2005 molecules near the BZ surface. This is expected as TIP4P2005 water molecules are a bit more strongly bounded to each other than SPC/E water molecules and near the less hydrophilic BZ surface, less water molecules are available for water--silica hydrogen bonds formation. This effect is not seen close to the GT surface and the water--silica hydrogen bonds are always higher close to the GT surface for the two water models than near the BZ surface. Thus, independly of the water model used, the GT model surface provides a strong hydrophilic surface. 